\documentclass[paper]{JHEP3}
\usepackage{epsfig}
\usepackage{slashed}
\usepackage{cite}
\def\bom#1{{\mbox{\boldmath $#1$}}}
\def\MSbar{$\overline{{\rm MS}}$}
\def\lapprox{\lower .7ex\hbox{$\;\stackrel{\textstyle <}{\sim}\;$}}
\def\gapprox{\lower .7ex\hbox{$\;\stackrel{\textstyle >}{\sim}\;$}}

\def\e{\epsilon}

\def\S{\, {\rm S}}

\def\CA{C_A}

\def\NF{N_F}

\def\sab{s_{12}}
\def\sac{s_{13}}
\def\sbc{s_{23}}

\def\sabc{s_{123}}

\def\Tbaa{T^\dagger_{211}}
\def\Tcaa{T^\dagger_{311}}
\def\Tbcb{T^\dagger_{232}}
\def\Tcab{T^\dagger_{312}}

\def\S{{\cal S}}
\def\eqref#1{(\ref{#1})}

\title{Two-Loop QCD Corrections to the Helicity Amplitudes for  $H \to$ 3~partons}

\author{T.~Gehrmann, M.~Jaquier\\Institut f\"ur Theoretische Physik, Universit\"at Z\"urich,
Winterthurerstrasse 190,\\CH-8057 Z\"urich, Switzerland }

\author{E.W.N.~Glover, A.~Koukoutsakis\\
IPPP, Department of Physics, University of Durham, Durham DH1 3LE, England}

\abstract{
Many search strategies for the Standard Model Higgs boson apply specific selection 
criteria on hadronic jets observed in association with the Higgs boson decay products, either 
in the form of a jet veto, or by defining event samples according to jet multiplicity. To improve
the theoretical description of Higgs-boson-plus-jet production (and the closely related Higgs 
boson transverse momentum distribution), we derive the two-loop QCD 
corrections to the helicity amplitudes for the processes $H \to
ggg$ and $H\to q \bar q g$ in an effective theory with
infinite top quark mass. The helicity amplitudes are extracted from
the coefficients appearing in the  general tensorial structure for each
process. The coefficients are  derived from the Feynman graph
amplitudes by means of projectors within the conventional dimensional
regularization scheme. The infrared pole structure of our result agrees with 
the expectation from infrared factorization and
the finite parts of  the amplitudes are expressed  in terms of one- and
two-dimensional harmonic polylogarithms.} 
\preprint{IPPP/11/77, ZU-TH 24/11}

\begin{document}
\unitlength1cm

\section{Introduction}
\setcounter{equation}{0}
\label{sec:intro}

Within the Standard Model of particle physics, the Higgs boson  is the only
particle remaining to be discovered. The Higgs boson is crucial for  electroweak
symmetry breaking, the  mechanism that explains the generation of the masses of
the fermions and the weak gauge bosons. While the vacuum expectation value of
the Higgs field is directly related to the Fermi constant, its mass remains a
free parameter that can be constrained but not predicted by the theory. 

The direct detection of the Higgs boson at LEP and the Tevatron has been a  very
challenging task over the past two decades~\cite{higgslep,higgstev}. The LEP
experiments~\cite{higgslep} excluded Higgs boson masses below $M_H\sim 114$~GeV,
while the  Tevatron excluded Higgs masses in a narrow window around the $W$-pair
threshold $M_H\sim 2M_W$.   With the start of the  Large Hadron Collider (LHC)
at CERN, the focus of the Higgs boson  searches has moved to the ATLAS and CMS
experiments, where the search is based on different decay channels. 
For $M_H\gapprox 135$~GeV
the decay into two weak gauge bosons is most prominent, while for lower Higgs 
boson masses
the search at the LHC is much more challenging,  
since the dominant decay modes are
overwhelmed by large Standard Model  backgrounds.  For example, the search for
light Higgs bosons with $M_H \lapprox 130$~GeV is based on the rare decay  $H\to
\gamma\gamma$, which has a branching ratio of  
$\mathcal{O}\left(10^{-3}\right)$, thus requiring larger integrated luminosity. 
For light Higgs boson masses, the dominant production
process at the LHC is gluon fusion.
 Based on  the first ${\cal O}$(5 fb$^{-1}$) of proton data taken
in 2011,   ATLAS~\cite{atlashiggs} and CMS~\cite{cmshiggs}  are now 
able to narrow
down  the  allowed mass range for the Standard Model Higgs boson considerably
by  essentially excluding the Higgs bosons in the range
 ${\cal O}(130~{\rm GeV})
\gapprox M_H \gapprox {\cal O}(600~{\rm GeV})$, while observing an excess 
of Higgs boson candidate events around $M_H=125$~GeV. 

At leading-order (LO), the Higgs coupling to the two gluons is mediated through
a quark loop. Since the Higgs coupling to the quarks is proportional to the
quark masses, the dominant contribution is generated from the top
quark~\cite{higgstop}.  The next-to-leading-order (NLO) corrections~\cite{spira}
to this process have  been calculated and turn out to be very large ($>$60\%).
In the heavy top quark limit, $M_{t} \rightarrow \infty$, the $Hgg$ coupling
becomes independent of $M_{t}$. One can therefore integrate out the top mass
($M_{t}$) and formulate an effective Langrangian $\mathcal{L}_{eff}$ for the
$Hgg$ coupling~\cite{effth}.  This technique is valid for $M_{H} < 2 M_{t}$ and
reduces the loops that need to be calculated by one.  In this limit, the
inclusive Higgs boson production cross section has been computed at
NLO~\cite{higgs1l} and  at next-to-next-to-leading-order
(NNLO)~\cite{higgsnnloinc}, indicating  a stabilization of the perturbative
prediction at this order.

Experimental searches of the Higgs boson apply final state cuts to improve the
significance of a potential signal over Standard Model background  processes. To
implement these cuts in the theoretical description,  fully exclusive
calculations, which keep track of the kinematical information  of all final
state particles (Higgs decay products and QCD radiation) are  mandatory. In the
heavy top quark limit, the NNLO corrections to Higgs  production via gluon
fusion have been computed fully exclusively, including the Higgs decay to two
photons or two weak gauge bosons by  two independent
groups~\cite{babishiggs,grazzinihiggs}. These calculations  are in the form of
flexible parton-level event generators, which can  properly account for the
final state restrictions used in the experimental  studies. 

An important final state discriminator is the number of jets observed  in
addition to the potential Higgs boson decay products, and the  Higgs signal can
often be enhanced by applying jet vetos~\cite{jetveto,t1}. In many searches, it
is however expected that the $H+0j$ and $H+1j$ samples  contribute roughly
equally to the sensitivity.   In the above-mentioned NNLO calculations, the
$H+1j$ final states are  included to NLO~\cite{h1j}, and the $H+2j$ final states
to LO. NLO corrections  to $H+2j$-production have been derived
recently~\cite{h2j}.  The correlation between  samples of different jet
multiplicity has recently  been matter of quite some   debate~\cite{t2,plehn},
and an improved theoretical description of  $H+1j$-production to NNLO accuracy
is essential in order to have  the  same   theoretical accuracy for the $H+1j$
contribution as for the $H+0j$  contribution. 

In the heavy top quark limit, a full NNLO QCD calculation 
of $H+1j$ production requires
the computation of the matrix elements of three contributions:
\begin{itemize}
\item[(a)] the
tree level $ H  \rightarrow 5$ partons amplitudes, 
\item[(b)]
 the
one-loop corrections to the $ H  \rightarrow 4$ partons
amplitudes,
\item[(c)]  the two-loop corrections to the $ H 
\rightarrow ggg$ and $ H  \rightarrow q\bar{q}g$  matrix elements.
\end{itemize}
The tree-level contributions of type (a) can be computed with standard 
tree-level methods, and compact expressions can be obtained by using
MHV-techniques~\cite{dixonglover}. The one-loop terms of type (b) 
were derived in an analytic form in~\cite{h1l}, and form part 
of the NLO corrections to $H+2j$ final states. The 
$H \rightarrow ggg$ and $ H  \rightarrow gq\bar{q}$  matrix elements 
were previously known to one loop~\cite{schmidt}. In this paper,
we compute the two-loop corrections to these processes in the heavy top quark limit. 
We note that expressions for the two-loop $H \to ggg$ helicity amplitudes were previously given in the PhD thesis of one of us.~\cite{thanos}

The three different contributions must be combined into a 
parton-level event generator program. All three are separately 
infrared-divergent, and only their sum is finite and physically 
meaningful. To combine the contributions, an infrared subtraction method is 
required. Several methods have been applied successfully in 
NNLO
calculations of exclusive observables in the recent past: 
sector decomposition~\cite{secdec}, $q_T$-subtraction~\cite{qtsub} and 
antenna subtraction~\cite{antsub,hadant,joao}. 
A resulting parton-level event generator
will allow an NNLO description of both $H+1j$ production and of the Higgs boson 
transverse momentum distribution. 

In addition to their phenomenological importance for precise predictions of 
collider processes, two-loop amplitudes are interesting to investigate 
fundamental properties of quantum field theory at high perturbative orders, aiming 
to identify regularities, asymptotic behaviour and heading towards an all-order 
understanding of field theory amplitudes. In this context, the massless $2\to 2$ QCD scattering 
amplitudes at two loops~\cite{twol} were used to determine the
high energy Regge trajectories of quarks and gluons 
at the two-loop level~\cite{regge}. Similarly,  the two-loop decay matrix elements 
for $\gamma^*\to q\bar q g$~\cite{3jme} and $H\to 3$~partons 
fix the the two-loop splitting amplitudes~\cite{2lsplit}, which describe 
the collinear factorization of loop amplitudes. 

This paper is structured as follows: in Section~\ref{sec:notation}, we consider
the effective coupling of the Higgs boson to light partons and define the
kinematics.  In Section~\ref{sec:gentensor},  we describe the method used to
construct the  tensor coefficients in the general amplitudes. In
Section~\ref{sec:helicity},  we construct the helicity amplitudes. The
derivation of the two-loop  corrections to the helicity amplitudes, their
renormalization and  infrared properties are described in
Section~\ref{sec:calc}. We conclude with  an outlook in Section~\ref{sec:conc}.
The  one-loop and two-loop helicity amplitudes  can be expressed in compact
analytic form, and are enclosed in the Appendix.

\section{Notation and kinematics}
\setcounter{equation}{0}
\label{sec:notation}

\subsection{The effective Lagrangian}

At tree level, the Higgs boson does not couple either to the gluon or to massless 
quarks.   In higher orders in perturbation theory, heavy quark loops introduce
a coupling between the Higgs boson and gluons. As we mentioned in Sec.~\ref{sec:intro}, in the heavy top quark limit, 
$M_{t}\rightarrow \infty$, the $Hgg$ coupling becomes
independent of $M_{t}$. We can therefore integrate out the top quark
field and formulate an effective Lagrangian, $\mathcal{L}_{eff}$ \cite{effth} 
that 
couples the scalar Higgs field and the gluon field strength tensor, thereby approximating the ${H}gg$
coupling.
This large top quark mass approximation has been shown to
work very well under the condition that the kinematic
scales involved are smaller
than twice the top quark mass~\cite{ptdist}.

The effective Lagrangian reads,
\begin{equation}
{\cal L}_{{\rm int}} = -\frac{\lambda}{4} H G_a^{\mu\nu} G_{a,\mu\nu}\ .
\label{eq:lagr}
\end{equation}
where $G_a^{\mu\nu}$ is the field strength tensor of the
gluon.
The coupling $\lambda$ has inverse mass dimension. It can be computed 
by matching~\cite{kniehl,kniehl2} 
the effective theory to the full standard 
model cross section~\cite{spira}.

\subsection{Kinematics}
\label{subsec:defs}
\setcounter{equation}{0}
We consider the decay of the Higgs boson to three gluons,
\begin{equation}
{{H}} (p_4) \longrightarrow g_1(p_1) + g_2(p_2) + g_3(p_3)\; , 
\end{equation}
or into a quark-antiquark pair and a gluon,
\begin{equation}
{{H}} (p_4) \longrightarrow q(p_1) + \bar q(p_2) + g_(p_3)\; .
\end{equation}
It is convenient to define the invariants,
\begin{equation}
\sab = (p_1+p_2)^2\;, \qquad \sac = (p_1+p_3)^2\;, \qquad 
\sbc = (p_2+p_3)^2\;,
\end{equation}
which fulfill 
\begin{equation}
p_{4}^2 = \sab + \sac + \sbc \equiv \sabc \equiv M_H^2\; ,
\end{equation}
as well as the dimensionless invariants, 
\begin{equation}
x = \sab/\sabc\;, \qquad y = \sac/\sabc\;, \qquad z = \sbc/\sabc\;,
\end{equation}
which satisfy $x+y+z=1$.

\section{The general tensors}
\label{sec:gentensor}

The  
amplitudes $|{\cal M}\rangle$ can be written as,
\begin{eqnarray}
\label{eq:Mdefhiggs}
|{\cal M}_{ggg}\rangle &=&S_{\mu \nu \rho}(g_{1};g_{2};g_{3})\epsilon_{1}^{\mu}\epsilon_{2}^{\nu}\epsilon_{3}^{\rho}\, ,\nonumber \\
|{\cal M}_{q\bar qg}\rangle &=&T_{\rho}(q,\bar q;g)\epsilon^{\rho}\, ,
\end{eqnarray}
while the partonic currents may be perturbatively decomposed as,
\begin{eqnarray}
S_{\mu \nu \rho}(g_{1};g_{2};g_{3}) &=& 
\lambda \sqrt{4\pi\alpha_s} f^{a_1a_2a_3}
\Big{[}S_{\mu \nu \rho}^{(0)}(g_{1};g_{2};g_{3})
+ \left(\frac{\alpha_s}{2\pi}\right)S_{\mu \nu \rho}^{(1)}(g_{1};g_{2};g_{3})\nonumber\\
&&+ \left(\frac{\alpha_s}{2\pi}\right)^2 S_{\mu \nu \rho}^{(2)}(g_{1};g_{2};g_{3}) 
+ {\cal O}(\alpha_s^3)\Big{]}, 
\label{eq:renormehggg}\\
T_{\rho}(q;\bar{q};g) &=& 
\lambda \sqrt{4\pi\alpha_s} T^a_{ij}
\Big{[}T_{\rho}^{(0)}(q;\bar{q};g)
+ \left(\frac{\alpha_s}{2\pi}\right) T_{\rho}^{(1)}(q;\bar{q};g)\nonumber\\
&&+ \left(\frac{\alpha_s}{2\pi}\right)^2 T_{\rho}^{(2)}(q;\bar{q};g)
+ {\cal O}(\alpha_s^3)\Big{]},
\label{eq:renormehqqg}
\end{eqnarray}
where $\alpha_s$ is the QCD coupling constant, 
and $S_{\mu \nu \rho}^{(i)}$ 
and $T_{\rho}^{(i)}$ are the $i$-loop
contributions to the amplitude.  The SU(3) generators are normalized as $tr(T^aT^b) = \delta^{ab}/2$.


\subsection{The general tensor for $H\to ggg$}
The most general tensor structure for the partonic current $\S_{\mu \nu \rho}(g_{1};g_{2};g_{3})$
is given by,
\begin{eqnarray}
\label{eq:gentensor}
S_{\mu \nu \rho}(g_{1};g_{2};g_{3})\epsilon_{1}^{\mu}\epsilon_{2}^{\nu}\epsilon_{3}^{\rho}
&=&\sum_{ i,j,k =1}^{3}\,A_{i\,j\,k}\,p_{i}\!\cdot\!\epsilon_{1}\,
p_{j}\!\cdot\!\epsilon_{2} \, p_{k}\!\cdot\!\epsilon_{3}
+\sum_{ i=1}^{3}\,B_{i}\,p_{i}\!\cdot\!\epsilon_{1}\,\epsilon_{2}\!\cdot\!\epsilon_{3}\nonumber\\ 
&+&\sum_{ i=1}^{3}\,C_{i}\,p_{i}\!\cdot\!\epsilon_{2}\,\epsilon_{1}\!\cdot\!\epsilon_{3}
+\sum_{ i=1}^{3}\,D_{i}\,p_{i}\!\cdot\!\epsilon_{3}\,\epsilon_{1}\!\cdot\!\epsilon_{2}\nonumber\\
 &=&A_{211}\,{ p_{2}\!\cdot\!\epsilon_{1}}\,{ p_{1}\!\cdot\!\epsilon_{2}}\,{ p_{1}\!\cdot\!\epsilon_{3}}+A_{212}\,{ p_{2}\!\cdot\!\epsilon_{1}}
\,{ p_{1}\!\cdot\!\epsilon_{2}}\,{ p_{2}\!\cdot\!\epsilon_{3}}+A_{231}\,{ p_{2}\!\cdot\!\epsilon_{1}}\,{ p_{3}\!\cdot\!\epsilon_{2}
}\,{ p_{1}\!\cdot\!\epsilon_{3}}\nonumber\\
&+&A_{232}\,{ p_{2}\!\cdot\!\epsilon_{1}}\,{ p_{3}\!\cdot\!\epsilon_{2}}\,{ 
p_{2}\!\cdot\!\epsilon_{3}}+A_{311}\,{ p_{3}\!\cdot\!\epsilon_{1}}\,{ p_{1}\!\cdot\!\epsilon_{2}}\,{
  p_{1}\!\cdot\!\epsilon_{3}}+A_{312}\,{ p_{3}\!\cdot\!\epsilon_{1}}\,{ p_{1}\!\cdot\!\epsilon_{2}}\,{ p_{2}\!\cdot\!\epsilon_{3}}\nonumber\\
&+&A_{331}\,{ p_{3}\!\cdot\!\epsilon_{1}}\,{ p_{3}\!\cdot\!\epsilon_{2}}\,{ p_{1}\!\cdot\!\epsilon_{3}}
+A_{332}\,{ p_{3}\!\cdot\!\epsilon_{1}}\,{ p_{3}\!\cdot\!\epsilon_{2}}\,{ p_{2}\!\cdot\!\epsilon_{3}}\nonumber\\
&+&B_{2}\,{ \epsilon_{2}\!\cdot\!\epsilon_{3}}\,{ p_{2}\!\cdot\!\epsilon_{1}}+B_{3}\,{\epsilon_{2}\!\cdot\!\epsilon_{3}}\,{p_{3}\!\cdot\!\epsilon_{1}}\nonumber\\
&+&C_{1}\,{\epsilon_{1}\!\cdot\!\epsilon_{3}}\,{ p_{1}\!\cdot\!\epsilon_{2}}+C_{3}\,{ \epsilon_{1}\!\cdot\!\epsilon_{3}
}\,{ p_{3}\!\cdot\!\epsilon_{2}}\nonumber\\
&+&D_{1}\,{ \epsilon_{1}\!\cdot\!\epsilon_{2}}\,{ p_{1}\!\cdot\!\epsilon_{3}}+D_{2}\,{ \epsilon_{1}\!\cdot\!\epsilon_{2}}\,{ 
p_{2}\!\cdot\!\epsilon_{3}}\, ,
\end{eqnarray}
where the constraints $p_{1} \cdot \epsilon_1 = 0$, $p_{2} \cdot \epsilon_2 =
0$ and $p_{3} \cdot \epsilon_3 = 0$ have been applied.
The tensor must satisfy the QCD Ward identity when the gluon polarization
vectors $\epsilon_1$, $\epsilon_2$ and $\epsilon_3$ are
replaced with the respective gluon momentum,
\begin{eqnarray}
(\epsilon_1 \to p_1)\rightarrow S_{\mu \nu \rho}(g_{1};g_{2};g_{3})p_{1}^{\mu}\epsilon_{2}^{\nu}\epsilon_{3}^{\rho} &=& 0\, ,\nonumber \\
(\epsilon_2 \to p_2)\rightarrow S_{\mu \nu \rho}(g_{1};g_{2};g_{3})\epsilon_{1}^{\mu}p_{2}^{\nu}\epsilon_{3}^{\rho} &=& 0\, ,\nonumber \\
(\epsilon_3 \to p_3)\rightarrow S_{\mu \nu \rho}(g_{1};g_{2};g_{3})\epsilon_{1}^{\mu}\epsilon_{2}^{\nu}p_{3}^{\rho} &=& 0\, .
\end{eqnarray}
These constraints yield relations amongst the 14 distinct tensor structures and 
applying
these identities give the gauge invariant form of the tensor,
\begin{eqnarray}\label{eq:HinAT}
S_{\mu \nu \rho}(g_{1};g_{2};g_{3})\epsilon_{1}^{\mu}\epsilon_{2}^{\nu}\epsilon_{3}^{\rho} &=& A_{211} T_{211} + A_{311} T_{311} + A_{232} T_{232}
+ A_{312} T_{312} \, ,
\end{eqnarray}
where $A_{ijk}$ are gauge independent functions and 
the tensor structures $T_{ijk}$ are given by,
\begin{eqnarray}\label{eq:Texpress}
T_{232}&=&{ p_{2}\!\cdot\! \epsilon_{1}}\,{ p_{3} \!\cdot\! \epsilon_{2}}\,{ p_{2} \!\cdot\! \epsilon_{3}}\!-\!\frac{1}{2}\,{
 \epsilon_{2} \!\cdot\! \epsilon_{3}}\,{ p_{2} \!\cdot\! \epsilon_{1}}\,{ \sbc}\!-\!{\frac {{ 
p_{3} \!\cdot\! \epsilon_{1}}\,{ p_{3} \!\cdot\! \epsilon_{2}}\,{ p_{2} \!\cdot\! \epsilon_{3}}\,{ \sab}}{{ \sac}}}\!+\!\frac{1}{2}\,{\frac {{ \epsilon_{2} \!\cdot\! \epsilon_{3}}\,{
 p_{3} \!\cdot\! \epsilon_{1}}\,{ \sbc}\,{ \sab}}{{ \sac}}} ,\nonumber \\
T_{211}&=&{ p_{2} \!\cdot\! \epsilon_{1}}\,{ p_{1} \!\cdot\! \epsilon_{2}}\,{ p_{1} \!\cdot\! \epsilon_{3}}\!-\!\frac{1}{2}\,{ 
\epsilon_{1} \!\cdot\! \epsilon_{2}}\,{ p_{1} \!\cdot\! \epsilon_{3}}\,{ \sab}\!-\!{\frac {{ p_{2} \!\cdot\! \epsilon_{1}}\,{ p_{1} \!\cdot\! \epsilon_{2}
}\,{ p_{2} \!\cdot\! \epsilon_{3}}\,{ \sac}}{{ \sbc}}}\!+\!\frac{1}{2}\,{\frac {{ \epsilon_{1} \!\cdot\! \epsilon_{2}}\,{
 p_{2} \!\cdot\! \epsilon_{3}}\,{ \sac}\,{ \sab}}{{ \sbc}}},\nonumber \\
T_{311}&=&{ 
p_{3} \!\cdot\! \epsilon_{1}}\,{ p_{1} \!\cdot\! \epsilon_{2}}\,{ p_{1} \!\cdot\! \epsilon_{3}}\!-\!\frac{1}{2}\,{ \epsilon_{1} \!\cdot\! \epsilon_{3}}\,{ p_{1} \!\cdot\! \epsilon_{2}}\,{ \sac}\!-\!{\frac {{ p_{3} \!\cdot\! \epsilon_{1}}
\,{ p_{3} \!\cdot\! \epsilon_{2}}\,{ p_{1} \!\cdot\! \epsilon_{3}}\,{ \sab}}{{ \sbc}}}\!+\!\frac{1}{2}\,{\frac {{ \epsilon_{1} \!\cdot\! \epsilon_{3}}\,{
 p_{3} \!\cdot\! \epsilon_{2}}\,{ \sac}\,{ \sab}}{{ \sbc}}},\nonumber \\
T_{312}&=&{ p_{3} \!\cdot\! \epsilon_{1}
}\,{ p_{1} \!\cdot\! \epsilon_{2}}\,{ p_{2} \!\cdot\! \epsilon_{3}}\!-\!{ p_{2} \!\cdot\! \epsilon_{1}}\,{ p_{3} \!\cdot\! \epsilon_{2}}\,{ p_{1} \!\cdot\! \epsilon_{3}}\!+\!\frac{1}{2}\,{ \epsilon_{1} \!\cdot\! \epsilon_{3}}\,{ p_{3} \!\cdot\! \epsilon_{2}}\,{ \sab}\!+\!\frac{1}{2}\,{ \epsilon_{1} \!\cdot\! \epsilon_{2}}\,{ p_{1} \!\cdot\! \epsilon_{3}}\,{
 \sbc}\nonumber\\
&&\!-\!\frac{1}{2}\,{ \epsilon_{1} \!\cdot\! \epsilon_{3}
}\,{ p_{1} \!\cdot\! \epsilon_{2}}\,{ \sbc}\!+\!\frac{1}{2}\,{ \epsilon_{2} \!\cdot\! \epsilon_{3}}\,{ p_{2} \!\cdot\! \epsilon_{1}}\,{ 
\sac}\!-\!\frac{1}{2}\,{ \epsilon_{1} \!\cdot\! \epsilon_{2}}\,{ p_{2} \!\cdot\! \epsilon_{3}}\,{ \sac}\!-\!\frac{1}{2}\,{ \epsilon_{2} \!\cdot\! \epsilon_{3}}\,
{ p_{3} \!\cdot\! \epsilon_{1}}\,{ \sab} \, .
\end{eqnarray}

The coefficients are functions of the invariants $\sab$, $\sbc$ and $\sac$ and are
further related by symmetry under the interchange of the three gluons,
\begin{eqnarray}
A_{211}(\sab,\sac,\sbc) &=& -A_{311}(\sac,\sab,\sbc)\, ,\nonumber \\
A_{232}(\sab,\sac,\sbc) &=& -A_{311}(\sab,\sbc,\sac)\, .
\end{eqnarray}

The coefficients $A_{ijk}$ may be easily extracted from a Feynman diagram calculation
using projectors such that,
\begin{equation}
\sum_{\rm spins} {\cal P}(A_{ijk})\,S_{\mu \nu \rho}(g_{1};g_{2};g_{3})\epsilon_{1}^{\mu}\epsilon_{2}^{\nu}\epsilon_{3}^{\rho} = A_{ijk}\, ,
\end{equation}
where the four projectors are given by,
\begin{eqnarray}
 {\cal P}( A_{311})&=&-{\frac { \left( D-4 \right) }{{
 \sab}\,{ \sbc}\,{ \sac}\, \left( D-3 \right) }}\Tbcb-{\frac {{ 
\sbc}\, \left( D-4 \right) }{{{ \sac}}^{2}{{ \sab}}^{
2} \left( D-3 \right) }}\Tbaa\nonumber\\
&&+{\frac {{ \sbc}\, D  }{{ \sab}\,{{ \sac}}^{3} \left( D-3 \right) }}\Tcaa-{\frac {
 \left( D-2 \right) }{{{ \sac}}^{2}{ \sab}\, \left( 
D-3 \right) }}\Tcab,\nonumber\\
 {\cal P}(A_{232})&=&{\frac {{ \sac}\, D  }{{ \sab}\,{{ \sbc}}^{3} \left( D-3 \right) }}\Tbcb+{
\frac { \left( D-4 \right) }{{ \sbc}\,{{ \sab}}^{2}
 \left( D-3 \right) }}\Tbaa\nonumber\\
&&-{\frac { \left( D-4 \right) }{{
 \sab}\,{ \sbc}\,{ \sac}\, \left( D-3 \right) }}\Tcaa+{\frac {
 \left( D-2 \right) }{{{ \sbc}}^{2}{ \sab}\, \left( 
D-3 \right) }}\Tcab,\nonumber\\
{\cal P}(A_{312})&=&{\frac { \left( D-2 \right) }{{{ \sbc}}^{2}{ \sab}\, \left( D-3 \right) }}\Tbcb+{\frac { \left(D-2 \right) }{{ \sac}\,{{ \sab}}^{2} \left( D-3
 \right) }}\Tbaa\nonumber\\
&&-{\frac { \left( D-2\right) }{{{ \sac}}^{2
}{ \sab}\, \left( D-3 \right) }}\Tcaa+{\frac {  D  }{{ \sab}\,{ \sbc}\,{ \sac}\, \left( D-3 \right) }}\Tcab,\nonumber\\
{\cal P}(A_{211})&=&{\frac { \left( D-4 \right) }{{ \sbc}\,{{
 \sab}}^{2} \left( D-3 \right) }}\Tbcb+{\frac {{ \sbc}\,  D
  }{{ \sac}\,{{ \sab}}^{3} \left( D-3
 \right) }}\Tbaa\nonumber\\
&&-{\frac {{ \sbc}\, \left( D-4 \right) }{{{
 \sac}}^{2}{{ \sab}}^{2} \left( D-3 \right) }}\Tcaa+{\frac { \left( D-2
 \right) }{{ \sac}\,{{ \sab}}^{2} \left( D-3
 \right) }}\Tcab . \nonumber\\
 \label{eq:Hgggprojectors}
\end{eqnarray}

Each of the tensor
coefficients $A_{ijk}$ has a perturbative expansion of the form,
\begin{eqnarray}
\label{eq:renorme}
A_{ijk} &=&  \lambda\sqrt{4\pi\alpha_s} \left[
A_{ijk}^{(0)}  
+ \left(\frac{\alpha_s}{2\pi}\right) A_{ijk}^{(1)}  
+ \left(\frac{\alpha_s}{2\pi}\right)^2 A_{ijk}^{(2)} 
+ {\cal O}\left((\alpha_s)^3\right) \right] \;, 
\end{eqnarray}
while the tree-level values are,
\begin{eqnarray}\label{eq:Atree}
A_{211}^{(0)} &=&\frac{2}{s_{13}} \; ,\nonumber \\
A_{311}^{(0)} &=&-\frac{2}{s_{12}} \; ,\nonumber \\
A_{232}^{(0)} &=&\frac{2}{s_{12}} \; ,\nonumber \\
A_{312}^{(0)} &=&-\frac{2}{s_{12}}-\frac{2}{s_{23}}-\frac{2}{s_{13}}\; .
\end{eqnarray}

\subsection{The general tensor for $H\to q\bar qg$}
The most general tensor structure for the partonic current $T_{\rho}(q;\bar q;g)$
is given by,
\begin{eqnarray}
T_{\rho}(q;\bar q;g) \epsilon_3^\rho & = & 
A_1 \bar u(p_1) \slashed p_3 v(p_2) p_1\cdot \epsilon_3 
+ A_2 \bar u(p_1) \slashed p_3 v(p_2) p_2\cdot \epsilon_3 
+ A_3 \bar u(p_1) \slashed \epsilon_3 v(p_2),
\label{eq:qtensor}
\end{eqnarray}
where $p_3\cdot \epsilon_3=0$ has been applied. The QCD Ward identity yields, 
$$ A_3 = -p_1\cdot p_3 A_1 - p_2\cdot p_3 A_2\,,$$
such that the amplitude can be written as, 
\begin{eqnarray}
T_{\rho}(q;\bar q;g) \epsilon_3^\rho & = &  A_1 \left( 
\bar u(p_1) \slashed p_3 v(p_2) p_2\cdot \epsilon_3 - \bar u(p_1) \slashed \epsilon_3 v(p_2) 
 p_2\cdot p_3 \right) \nonumber \\
 & & + A_2  \left( 
\bar u(p_1) \slashed p_3 v(p_2) p_1\cdot \epsilon_3 - \bar u(p_1) \slashed \epsilon_3 v(p_2) 
 p_1\cdot p_3 \right) \label{eq:qtensfull} \\
 & \equiv & A_1 T_1 + A_2 T_2\;.
\end{eqnarray}

The coefficients $A_i$ can be extracted from a Feynman diagram calculation 
by using projectors such that,
\begin{equation}
\sum_{{\rm spins}} {\cal P} (A_i) T_{\rho}(q;\bar q;g) \epsilon_3^\rho = A_i\,,
\end{equation}
where the projectors are given by,
\begin{eqnarray}
{\cal P} (A_1) &=& \frac{(D-2)}{2(D-3)s_{12}s_{13}^2} T^\dagger_1 
-\frac{(D-4)}{2(D-3)s_{12}s_{13}s_{23}}T^\dagger_2 \,, \\
{\cal P} (A_2) &=&  -\frac{(D-4)}{2(D-3)s_{12}s_{13}s_{23}}T^\dagger_1 
+\frac{(D-2)}{2(D-3)s_{12}s_{23}^2}T^\dagger_2 \,.
 \label{eq:Hqqgprojectors}
\end{eqnarray}

Each of the 
coefficients $A_{i}$ has a perturbative expansion of the form,
\begin{eqnarray}
\label{eq:renormeq}
A_{i} &=&  \lambda\sqrt{4\pi\alpha_s}\left[
A_{i}^{(0)}  
+ \left(\frac{\alpha_s}{2\pi}\right) A_{i}^{(1)}  
+ \left(\frac{\alpha_s}{2\pi}\right)^2 A_{i}^{(2)} 
+ {\cal O}\left((\alpha_s)^3\right) \right] \;, 
\end{eqnarray}
while the tree-level values are simply,
\begin{equation}\label{eq:Aqtree}
A_{1}^{(0)} =A_2^{(0)} = \frac{1}{s_{12}}\; .
\end{equation}

\section{Helicity amplitudes}
\label{sec:helicity}
\setcounter{equation}{0}

The general form of the renormalized helicity amplitude
$|{\cal{M}}_{ggg}^{\lambda_{1}\lambda_{2}\lambda_{3}}\rangle$ for the decay,
$
H(p_{4})\rightarrow g_{1}(p_{1},\lambda_{1})+g_{2}(p_{2},\lambda_{2})+g_{3}(p_{3},\lambda_{3})
$ can be written as,
\begin{eqnarray}
|{\cal{M}}_{ggg}^{\lambda_{1}\lambda_{2}\lambda_{3}}\rangle = S_{\mu \nu \rho}(g_{1};g_{2};g_{3})\epsilon^{\mu}_{1,\lambda_{1}}(p_1)\epsilon^{\nu}_{2,\lambda_{2}}(p_2)\epsilon^{\rho}_{3,\lambda_{3}}(p_3)\, ,
\end{eqnarray}
where the $\lambda_{i}=\pm$ denote helicity.
Similarly, the amplitude for the decay 
$|{\cal{M}}_{q\bar qg}^{\lambda_{1}\lambda_{2}\lambda_{3}}\rangle$ for the decay,
$
{{H}}(p_{4})\rightarrow q(p_{1},\lambda_{1})+\bar q(p_{2},\lambda_{2})+g(p_{3},\lambda_{3})
$ can be written as,
\begin{eqnarray}
|{\cal{M}}_{q\bar q g}^{\lambda_{1}\lambda_{2}\lambda_{3}}\rangle = T_{\rho}(q^{\lambda_1};\bar q^{\lambda_2};g)\epsilon^{\rho}_{3,\lambda_{3}}(p_3)\, .
\end{eqnarray}

The helicity amplitudes 
can be obtained from the general $D$-dimensional tensors of
Eqs.~\eqref{eq:gentensor} and \eqref{eq:qtensor} by setting 
the dimensionality of the Lorentz matrices
to be four and 
using standard four-dimensional helicity techniques 
\cite{Xu:1986xb,Berends:ie,dixon}.  This  corresponds to working in
the 't Hooft-Veltman scheme.  We use the standard convention of denoting
the two helicity states of a
four-dimensional light-like spinor $\psi(p)$  by,
\begin{equation}
\psi_\pm(p)= \frac{1}{2}(1\pm \gamma_5)\psi(p),
\end{equation}
with the further notation,
\begin{equation}
|p \pm \rangle = \psi_\pm(p), \qquad\qquad
\langle p\! \pm \!| = \overline{\psi}_\pm(p).
\end{equation}
Particles may thus be crossed to the initial state by reversing the sign of the
helicity.
The basic quantity is the spinor product,
\begin{equation}
\langle pq\rangle = \langle p\! - \!| q + \rangle, \qquad\qquad
[ pq ]= [ p\! + \!| q - ],
\end{equation}
such that 
\begin{equation}
\langle pq\rangle  [qp] = 2 p\cdot q.
\end{equation}
The polarization vector of a outgoing light-like particle 
with momentum $p$ can then be written
as
\begin{equation}
\epsilon_\pm^\mu(p;q) = \pm \frac{\langle q\! \mp \!| \gamma^\mu | p \mp
  \rangle}{\sqrt{2}\langle q\! \mp \!| p \pm \rangle} 
\end{equation}
where $q$ is a light-like reference momentum that satisfies $q\cdot p
\neq 0$ but which otherwise can be chosen freely.

Important identities relating spinorial objects are the
Fierz rearrangement,
\begin{equation}
  \langle p\!+ \!| \gamma^\mu \!| q+\rangle \langle r\!+\! |\gamma^\mu | s+
  \rangle = 2 [pr]\langle sq \rangle
\end{equation}
and charge conjugation,
\begin{equation}
  \langle p\!+\! | \gamma^\mu | q+\rangle = \langle q\!-\! | \gamma^\mu |p-
  \rangle .
\end{equation}

Substituting Eq.~(\ref{eq:Texpress}) into Eq.~(\ref{eq:HinAT}),  we can express
the helicity 
amplitudes 
for $H\to ggg$ directly in terms of spinor products. It turns out that the only two
 independent helicity amplitudes are $|{\cal M}_{ggg}^{+++}\rangle$ and
 $|{\cal M}_{ggg}^{++-} \rangle$. 
The other helicity amplitudes are obtained from $|{\cal{M}}_{ggg}^{+++}\rangle$ and
$|{\cal{M}}_{ggg}^{++-}\rangle$ by the usual parity relation and by exploiting the symmetry of the gluons.
Explicitly, choosing $p_{i+1}$ as reference momentum for $\epsilon_{i,\lambda_i}$ 
we find,
\begin{eqnarray}
|{\cal{M}}_{ggg}^{+++}\rangle &=& \alpha\, \frac{1}{\sqrt{2}} \frac{M_H^4}{
\langle p_{1} p_{2}\rangle \langle p_{2} p_{3}\rangle
 \langle p_{3} p_{1}\rangle}\nonumber ,\\
|{\cal{M}}_{ggg}^{++-}\rangle&=&\beta\, \frac{1}{\sqrt{2}}  \frac{[p_{1}
p_{2}]^{3}}{[p_{2} p_{3}][p_{1} p_{3}]}, 
\label{eq:hggghel}
\end{eqnarray}
where the coefficients $\alpha$ and $\beta$ are written 
in terms of the tensor coefficients,
\begin{eqnarray}
\label{eq:alphabeta}
\alpha &=&\frac{s_{12}s_{13}s_{23}}{2M_H^4}\left( \frac{\sab}{\sbc} A_{211}+ \frac{\sbc}{\sac} A_{232}-\frac{\sac}{\sbc} A_{311}-2 A_{312}\right),  \nonumber \\
\beta &=& \frac{\sac}{2} A_{211}. 
\end{eqnarray}

Likewise (\ref{eq:qtensfull}) yields the helicity amplitudes for 
$H\to q\bar q g$ in terms of spinor products. There is only one independent
helicity amplitude $|{\cal M}_{q\bar qg}^{-++} \rangle$ and all other amplitudes can be obtained 
from $|{\cal M}_{q\bar qg}^{-++} \rangle$ using the usual parity and 
charge conjugation relations. By choosing $p_1$ as reference momentum for 
 $\epsilon_{3,\lambda_3}$, we obtain,
\begin{eqnarray}
|{\cal{M}}_{q\bar q g}^{-++}\rangle &=& \gamma\, \frac{1}{\sqrt{2}} 
\frac{[p_{2} p_{3}]^2}{[p_1 p_2]}\, .
\label{eq:hqqghel}
\end{eqnarray}
The helicity coefficient $\gamma$ is obtained from the tensor 
coefficients as,
\begin{equation}
\label{eq:gamma}
\gamma = s_{12}\, A_1\;.
\end{equation}

As with the tensor coefficients, the  
helicity amplitude coefficients $\alpha$, $\beta$ and $\gamma$ are 
vectors in colour space and
have perturbative expansions,
\begin{equation}
\Omega =  \lambda \sqrt{4\pi\alpha_s}T_\Omega \left[
\Omega^{(0)}  
+ \left(\frac{\alpha_s}{2\pi}\right) \Omega^{(1)}  
+ \left(\frac{\alpha_s}{2\pi}\right)^2 \Omega^{(2)} 
+ {\cal O}(\alpha_s^3) \right] \;,\nonumber \\
\label{eq:omegaren}
\end{equation}
for $\Omega = \alpha,\beta,\gamma$. The colour factor is 
$T_{\alpha}=T_{\beta} = f^{a_1a_2a_3}$ and $T_\gamma = T^{a_3}_{i_1j_2}$.

\section{Calculation of the two-loop helicity coefficients}
\label{sec:calc}
\setcounter{equation}{0}

\subsection{Calculation of two-loop Feynman amplitudes}
The calculation of the two-loop Feynman amplitudes contributing to 
$H\to ggg$ and $H\to q\bar qg$ follows closely the calculation of the 
two-loop helicity amplitudes for $\gamma^*\to q\bar q g$~\cite{3jme}, 
which contribute to the NNLO corrections to $e^+e^-\to 3j$ and 
related event shapes~\cite{ourevent,weinzierl}, and 
of the two-loop helicity amplitudes 
for $q\bar q \to V\gamma$~\cite{tancredi}. We performed two completely 
independent calculations of the amplitudes, which provide a 
strong internal cross-check on the results. 

The Feynman diagrams  contributing to the $i$-loop 
amplitude $|{\cal M}^{(i)}\rangle$ ($i=0,1,2$) were all generated using 
QGRAF~\cite{qgraf}. For $H\to ggg$, 
there are four diagrams at tree-level, 
60 diagrams 
at one loop and 1306 diagrams at two loops,
while for $H\to q\bar q g$, we have one diagram at tree-level, 
15 diagrams 
at one loop and 228 diagrams at two loops. 
We use dimensional 
regularization~\cite{dreg1,dreg2,hv} with $D=4-2\e$ dimensions.
 We therefore apply the $D$-dimensional projectors given in Eqs.~\eqref{eq:Hgggprojectors} and \eqref{eq:Hqqgprojectors}
and perform the summation over colours and spins using computer algebra
methods, mainly implemented in FORM~\cite{form}. 
When summing over the polarizations of the external 
gluons in the projectors, 
we use the axial gauge with a $D$-dimensional metric. 
Internal gluons are kept in Feynman 
gauge, resulting in internal ghost contributions to the loop amplitudes. 
The integrals appearing in the individual  two-loop
diagrams contain up to seven
propagators  in the denominator, and up to five irreducible scalar products in
the  numerator (i.e.\ scalar products which can not be expressed 
as linear combinations of the occurring propagators). 

The reduction of the two-loop integrals to 
a small set of master integrals using 
integration-by-parts (IBP)~\cite{chet1,chet2} and Lorentz invariance
(LI)~\cite{gr} identities was  performed
using the Laporta algorithm~\cite{laporta}, which is 
based on a lexicographic ordering of the integrals. We used  
two independent implementations of the Laporta algorithm: 
the MAPLE and FORM based implementation which was 
developed in the context of~\cite{3jme} and the 
recently developed C++ code 
 REDUZE~\cite{reduze}. Both implementations are based on auxiliary 
topologies~\cite{3jme}, and substantial work was required to automate 
the translation of the momentum assignments in the diagrams generated by 
QGRAF into the momentum sets of the auxiliary topologies. This process has 
been automated in FORM using an iterated shifting and matching algorithm for 
the momenta.

The two-loop master integrals 
relevant for this calculation are 
two-loop four-point functions with one leg off-shell.  These
functions were all computed in~\cite{mi} in  dimensional
regularization.  The results of~\cite{mi}
take the form of a Laurent series in $\e$, starting  at $\e^{-4}$, with
coefficients containing one- and two-dimensional  harmonic
polylogarithms (HPLs~\cite{hpl} and 2dHPLs~\cite{mi}), 
which are a generalization of  Nielsen's
polylogarithms~\cite{nielsen}. Several numerical implementations 
of HPLs and 2dHPLs are available~\cite{hplnum}.

Inserting the master integrals into the amplitudes and truncating 
the Laurent series to the required order, the unrenormalized one-loop and 
two-loop helicity coefficients are obtained. Their 
Laurent expansion contains HPLs and 2dHPLs up to weight 4. 
The 
expressions for the master integrals derived in~\cite{mi} apply to 
the kinematical situation of a $1\to 3$ decay, while the $H+1j$ production 
corresponds to a $2\to 2$ scattering process which is obtained from the 
decay kinematics by crossing. The crossing of the amplitudes requires the 
analytic continuation of the master integrals, which is described in 
detail in~\cite{ancont}.

\subsection{Ultraviolet renormalization}

Renormalization of
ultraviolet divergences is  performed in the $\overline{{\rm MS}}$
scheme.
It is carried out by replacing 
the bare coupling $\alpha_0$ with the renormalized coupling 
$\alpha_s\equiv \alpha_s(\mu^2)$,
evaluated at the renormalization scale $\mu^2$,
\begin{equation}
\alpha_0\mu_0^{2\e} S_\e = \alpha_s \mu^{2\e}\left[
1- \frac{\beta_0}{\e}\left(\frac{\alpha_s}{2\pi}\right) 
+\left(\frac{\beta_0^2}{\e^2}-\frac{\beta_1}{2\e}\right)
\left(\frac{\alpha_s}{2\pi}\right)^2+{\cal O}(\alpha_s^3) \right]\; ,
\end{equation}
where
\begin{displaymath}
S_\e =(4\pi)^\e e^{-\e\gamma}\qquad \mbox{with Euler constant }
\gamma = 0.5772\ldots
\end{displaymath}
and $\mu_0^2$ is the mass parameter introduced 
in dimensional regularization~\cite{dreg1,dreg2,hv} to maintain a 
dimensionless coupling 
in the bare QCD Lagrangian density; $\beta_0$ and $\beta_1$ are the first 
two coefficients of the QCD $\beta$-function,
\begin{equation}
\beta_0 = \frac{11 \CA - 4 T_R \NF}{6},  \qquad 
\beta_1 = \frac{17 \CA^2 - 10 C_A T_R \NF- 6C_F T_R \NF}{6}\;,
\end{equation}
with the QCD colour factors,
\begin{equation}
\CA = N,\qquad C_F = \frac{N^2-1}{2N},
\qquad T_R = \frac{1}{2}\; .
\end{equation}

The renormalization relation for the effective coupling $\lambda$ is given 
in~\cite{Harlander:2001is} as,
\begin{equation}
\lambda^{U}=\lambda\left[
1- \frac{\beta_0}{\e}\left(\frac{\alpha_s}{2\pi}\right) 
+\left(\frac{\beta_0^2}{\e^2}-\frac{\beta_1}{\e}\right)
\left(\frac{\alpha_s}{2\pi}\right)^2+{\cal O}(\alpha_s^3) \right]\; .
\end{equation}

We denote the $i$-loop contribution to the unrenormalized coefficients by 
$\Omega^{(i),{\rm U}}$, using the same normalization as 
for the decomposition of the renormalized amplitude (\ref{eq:omegaren}).
The renormalized coefficients are then obtained as,
\begin{eqnarray}
\Omega^{(0)}  &=& \Omega^{(0),{\rm U}} ,
 \nonumber \\
\Omega^{(1)}  &=& 
S_\e^{-1} \Omega^{(1),{\rm U}} 
-\frac{3\beta_0}{2\e} \Omega^{(0),{\rm U}}  ,  \nonumber \\
\Omega^{(2)} &=& 
S_\e^{-2} \Omega^{(2),{\rm U}}  
-\frac{5\beta_0}{2\e} S_\e^{-1}
\Omega^{(1),{\rm U}}  
-\left(\frac{5\beta_1}{4\e}-\frac{15\beta_0^2}{8\e^2}\right)
\Omega^{(0),{\rm U}}.
\end{eqnarray}

For the remainder of this paper we will set the renormalization scale
$\mu^2 = M_H^2 = s_{123}$. 
The full scale dependence of the helicity coefficients are given by,
\begin{eqnarray}
\Omega &=& \lambda \sqrt{4\pi\alpha_s(\mu^2)}T_\Omega \bigg\{
\Omega^{(0)}  + \left(\frac{\alpha_s(\mu^2)}{2\pi}\right)
\left[
\Omega^{(1)}  
+\frac{3}{2} \beta_0 \Omega^{(0)} \ln\left({\mu^2\over s_{123}}\right)
\right]
\nonumber \\
&&+ \left(\frac{\alpha_s(\mu^2)}{2\pi}\right)^2
\bigg[\Omega^{(2)} 
+\biggl(\frac{5}{2}\beta_0\Omega^{(1)} +\frac{5}{2}\beta_1\Omega^{(0)} \biggr) 
\ln\left({\mu^2\over s_{123}}\right)  
+\frac{15}{8}\beta_0^2 \Omega^{(0)}  
\ln^2\left({\mu^2\over s_{123}}\right)\bigg] \nonumber \\ &&
 + {\cal O}(\alpha_s^3) \bigg\}.  
\end{eqnarray}

\subsection{Infrared factorization}

The amplitudes contain infrared singularities that will be  analytically
canceled by those occurring in radiative processes of the
same order (ultraviolet divergences are removed by renormalization).
Catani~\cite{catani} has shown how to organize the 
infrared pole structure of the one- and two-loop contributions renormalized in the 
\MSbar\ scheme in terms of the tree and renormalized one-loop amplitudes.
This formula for the pole structure is proven~\cite{cataniproof}  from the 
structure of soft and collinear radiation in perturbation 
theory and can be generalized to higher loop order. 

The same factorization of pole terms applies to the 
helicity coefficients. 
In particular, the infrared behaviour of the one-loop coefficients is given by,
\begin{eqnarray}
\Omega^{(1)} &=& {\bom I}_\Omega^{(1)}(\epsilon) 
\Omega^{(0)} + \Omega^{(1),finite},
\end{eqnarray}
while the two-loop singularity structure is,
\begin{eqnarray}
\label{eq:polesa}
\Omega^{(2)} &=& \Biggl (-\frac{1}{2}  {\bom I}_\Omega^{(1)}(\epsilon) 
{\bom I}_\Omega^{(1)}(\epsilon)
-\frac{\beta_0}{\epsilon} {\bom I}_\Omega^{(1)}(\epsilon)
\nonumber \\ && 
+e^{-\epsilon \gamma } \frac{ \Gamma(1-2\epsilon)}{\Gamma(1-\epsilon)} 
\left(\frac{\beta_0}{\epsilon} + K\right)
{\bom I}_\Omega^{(1)}(2\epsilon) + {\bom H}_\Omega^{(2)}(\epsilon) 
\Biggr )\Omega^{(0)}\nonumber \\
&& + {\bom I}_\Omega^{(1)}(\epsilon) \Omega^{(1)}+ \Omega^{(2),finite},
\end{eqnarray}
where the constant $K$ is,
\begin{equation}
K = \left( \frac{67}{18} - \frac{\pi^2}{6} \right) \CA - 
\frac{10}{9} T_R \NF.
\end{equation}

For each of the processes under consideration, there is 
only one colour structure present at
tree level.
Adding higher loops does not
introduce additional colour structures, and the amplitudes are therefore
vectors in a one-dimensional space.  Similarly, 
the infrared 
singularity operators $\bom{I}_\Omega^{(1)}(\epsilon)$ are  $1 \times 1$ 
matrices in the colour space
and are given by,
\begin{eqnarray}
\bom{I}_\alpha^{(1)}(\epsilon)
& =& 
- \frac{e^{\epsilon\gamma}}{2\Gamma(1-\epsilon)} \Biggl[
N \left(\frac{1}{\epsilon^2}+\frac{\beta_0}{N\epsilon}\right) 
\left({\tt S}_{12}+{\tt S}_{13}+{\tt S}_{23}\right)\Biggr ]\; ,\label{eq:I1a}\\
&=& \bom{I}_\beta^{(1)}(\epsilon),\\
\bom{I}_\gamma^{(1)}(\epsilon)
& =& 
- \frac{e^{\epsilon\gamma}}{2\Gamma(1-\epsilon)} \Biggl[
N \left(\frac{1}{\epsilon^2}+\frac{3}{4\epsilon}+\frac{\beta_0}{2N\epsilon}\right) 
\left({\tt S}_{13}+{\tt S}_{23}\right)-\frac{1}{N}
\left(\frac{1}{\epsilon^2}+\frac{3}{2\epsilon}\right)
{\tt S}_{12}\Biggr ]\; ,\label{eq:I1c}
\end{eqnarray}
where, since we have set $\mu^2 = s_{123}$,
\begin{equation}
{\tt S}_{ij} = \left(-\frac{s_{123}}{s_{ij}}\right)^{\epsilon}.
\end{equation}
Note that on expanding ${\tt S}_{ij}$,
imaginary parts are generated, the sign of which is fixed by the small imaginary
part $+i0$ of $s_{ij}$.
The origin of the various terms in Eqs.~(\ref{eq:I1a})--(\ref{eq:I1c}) 
is straightforward.  Each parton pair $ij$
in the event forms a radiating antenna of scale $s_{ij}$.  
Terms proportional to ${\tt S}_{ij}$ are canceled by real radiation emitted from leg
$i$ and absorbed by leg $j$. The soft singularities ${\cal O}(1/\epsilon^2)$ are independent of
the identity of the participating partons and are universal.
However, the collinear singularities depend on the identities of the participating partons.  For
each quark we find a contribution of $3/(4\epsilon)$ and for each gluon we find a contribution
of $\beta_0/(2\epsilon)$ coming from the integral over the collinear splitting function.

Finally, the last term of Eq.~(\ref{eq:polesa}) that involves 
${\bom H}^{(2)}(\epsilon)$ 
produces only a single pole in $\epsilon$ and is given by,
\begin{equation}
\label{eq:htwo}
{\bom H}_\Omega^{(2)}(\epsilon)
=\frac{e^{\epsilon \gamma}}{4\,\epsilon\,\Gamma(1-\epsilon)} H_\Omega^{(2)} \;,  
\end{equation}
where the constant $H_\Omega^{(2)}$ is renormalization scheme dependent.
As with the single pole parts of $\bom{I}_\Omega^{(1)}(\epsilon)$,
the process-dependent
$H_\Omega^{(2)}$ can be constructed by counting the number of
radiating partons present in the event.
In our case, 
\begin{eqnarray}
H_\alpha^{(2)} = H_\beta^{(2)}&=&  3H^{(2)}_g,\nonumber \\
H_\gamma^{(2)} &=& 2H^{(2)}_q+H^{(2)}_g,
\end{eqnarray}
where, in the \MSbar\ scheme,
\begin{eqnarray}
H^{(2)}_q &=&
\left({7\over 4}\zeta_3+{\frac {409}{864}}- {\frac {11\pi^2}{96}}
\right)N^2
+\left(-{1\over 4}\zeta_3-{41\over 108}-{\pi^2\over 96}\right)
+\left(-{3\over 2}\zeta_3-{3\over 32}+{\pi^2\over 8}\right){1\over
N^2}\nonumber \\
&&
+\left({\pi^2\over 48}-{25\over 216}\right){(N^2-1)N_F\over N}\;, \\
H^{(2)}_g &=&  
\left(\frac{1}{2}\zeta_3+{\frac {5}{12}}+ {\frac {11\pi^2}{144}}
\right)N^2
+{\frac {5}{27}}\,\NF^2
+\left (-{\frac {{\pi }^{2}}{72}}-{\frac {89}{108}}\right ) N \NF 
-\frac{\NF}{4N}. 
\end{eqnarray}

At leading order, one can insert the values of the tensorial coefficients given in
Eqs.~\eqref{eq:Atree} and \eqref{eq:Aqtree}, into Eqs.~\eqref{eq:alphabeta} and \eqref{eq:gamma} respectively to find,
\begin{equation}
\alpha^{(0)}
= \beta^{(0)} =   \gamma^{(0)} = 1\, .
\end{equation}
The renormalized NLO helicity amplitude coefficients can be 
straightforwardly obtained to all orders in $\epsilon$ from 
the  helicity coefficients $\Omega^{(1)}$.   
For practical purposes they are needed through to ${\cal O}(\epsilon^2)$  
in evaluating the one-loop self-interference and the 
infrared divergent one-loop contribution to the two-loop amplitude, while
only the finite piece is needed for the one-loop self-interference.
They can be decomposed 
according to their colour structure as follows,
\begin{eqnarray}\label{eq:wm1}
\Omega^{(1),finite} &=& \Biggl (
N\ A^{(1)}_\Omega + \frac{1}{N}\ B^{(1)}_\Omega + \NF C^{(1)}_\Omega \Biggr
).
\end{eqnarray} 

The finite two-loop remainder is  obtained by subtracting the
predicted infrared structure (expanded through to ${\cal O}(\epsilon^0)$) from
the renormalized helicity coefficient.  We further decompose the 
finite remainder according to the colour casimirs as follows,
\begin{eqnarray}\label{eq:wm2}
\Omega^{(2),finite} &=&  \Biggl (
N^2 A^{(2)}_\Omega +N^0 B^{(2)}_\Omega +\frac{1}{N^2} C^{(2)}_\Omega +\frac{\NF}{N} D^{(2)}_\Omega +  N\NF E^{(2)}_\Omega + 
\NF^2 F^{(2)}_\Omega \Biggr
). \nonumber \\
\end{eqnarray}
All one- and two-loop coefficients are given in
Appendix A and B respectively.

To calculate the two-loop contributions to Higgs-boson-plus-jet production 
at hadron colliders, the helicity amplitudes must be crossed to the 
appropriate kinematical situations. Two types crossings are required:
\begin{eqnarray}
&&g(p_1) + g(p_2) \to H(p_4) + g(-p_3)\;,\qquad q(p_1) + \bar q(p_2) \to H(p_4) 
+ g(-p_3)\;,  \\
&&g(p_2) + g(p_3) \to H(p_4) + g(-p_1)\;, 
\qquad \bar q(p_2)+ g(p_3) \to H(p_4) 
+ \bar q(-p_1)\;.
\end{eqnarray}
The definitions of the helicity amplitudes in terms of 
momentum spinors (\ref{eq:hggghel},\ref{eq:hqqghel}) remain 
unchanged by the crossing, such that only the helicity 
coefficients $\alpha,\beta,\gamma$ are to be continued to the 
appropriate kinematical region. The analytical continuation 
of the polylogarithmic functions appearing in two-loop 
amplitudes is described in detail in~\cite{ancont,tancredi}. We provide the 
one-loop and two-loop coefficients in all relevant analytic continuations 
in FORM format with the arXiv-submission 
of this article.

\section{Conclusions}
\label{sec:conc}
\setcounter{equation}{0}
In this paper, we derived the two-loop corrections to the helicity amplitudes 
for the processes $H\to ggg$ and $H\to q\bar qg$. Our calculation was 
performed in dimensional regularization by applying $D$-dimensional 
projection operators to the most general tensor structure of the amplitude. 
 Our results 
are expressed in terms of dimensionless helicity coefficients, which multiply 
the basic tree-level amplitudes, expressed in four-dimensional 
spinors. 
 By applying Catani's infrared factorization 
formula, we extract the finite parts of the helicity coefficients, which 
are independent on the precise scheme used to define the helicity amplitudes. 
We provide compact analytic expressions for the two-loop helicity coefficients 
in terms of HPLs and 2dHPLs. 

By crossing the 
Higgs boson to the final state and two partons to the initial state, these 
amplitudes describe the two-loop corrections to the parton-level process for 
$H+1j$ production and for the transverse momentum distribution of the Higgs 
boson. To compute the NNLO corrections to these processes, the newly 
derived two-loop amplitudes need to be combined with the previously 
known~\cite{dixonglover,h1l} 
crossings of the tree-level amplitudes for $H\to 5$~partons and 
one-loop amplitudes for $H\to 4$~partons into a parton-level event generator. 
The tree-level double-real radiation and one-loop real-virtual contributions both contain 
infrared singularities from soft or collinear real radiation. In order 
to numerically implement these contributions, the singular contributions 
must be subtracted and combined with infrared singularities from the two-loop integrals. Up to now, infrared subtraction at NNLO has not been 
fully
accomplished for hadron collider processes involving final state jets.
In the context of dijet production, a first proof-of-principle 
implementation of the double-real radiation contribution  
exists~\cite{joao}, and the infrared structure of the 
NNLO subtraction 
terms for hadron collider processes is largely understood~\cite{hadant}.  
With the 
amplitudes derived in this paper, it should thus become feasible to 
compute the NNLO corrections 
to  $H\to 1j$ production and the Higgs boson transverse momentum 
distribution at hadron colliders.

\section*{Acknowledgments} 
This work has been supported in part by the Forschungskredit der 
Universit\"at Z\"urich, by the Swiss National 
 Science Foundation (SNF) under contract
200020-138206 and  
 by the Research Executive Agency (REA) of the
European Union under the Grant Agreement number PITN-GA-2010-264564
(LHCPhenoNet).
EWNG gratefully
acknowledges the support of the Wolfson Foundation and the Royal Society
 and thanks the Institute for Theoretical Physics at the ETH for
 its kind hospitality during the completion of this work.

\begin{appendix}
\renewcommand{\theequation}{\mbox{\Alph{section}.\arabic{equation}}}

\section{One-loop helicity coefficients}
\setcounter{equation}{0}
The finite contributions to the 
renormalized one-loop helicity coefficients, decomposed 
in colour factors according to (\ref{eq:wm1}) are:
\begin{eqnarray}
A^{(1)}_\alpha & = & \bigg[\frac{1}{2}\Big(-G(1-z,0,y)-H(1,0,z)-G(0,1-z,y)\nonumber\\
&&-\:H(0,1,z)-H(0,z)G(1-z,y)+G(0,y)H(1,z)-G(0,y)H(0,z)\Big)\nonumber\\
&&\qquad+\:G(-z,1-z,y)+G(1,0,y)-H(1,z)G(-z,y)\nonumber\\
&&\qquad+\:\frac{11}{12}\Big(-G(1-z,y)+H(1,z)-H(0,z)-G(0,y)\Big)-\frac{\pi^2}{12}
\bigg]\nonumber\\
&&+\:\frac{1}{6}\Big((y+z)(1-y)-z^2)+i\frac{11\pi}{4}\;,\\
B^{(1)}_\alpha & = & 0\;, \\
C^{(1)}_\alpha & = & \frac{1}{6}\bigg[G(1-z,y)-H(1,z)+H(0,z)+G(0,y)\bigg]\nonumber\\
&&-\:\frac{1}{6}\Big((y+z)(1-y)-z^2\Big)-i\frac{\pi}{2}\;,\\
A^{(1)}_\beta & = & \bigg[\frac{1}{2}\Big(-G(1-z,0,y)-H(1,0,z)-G(0,1-z,y)-H(0,1,z)\nonumber\\
&&-\:H(0,z)G(1-z,y)+G(0,y)H(1,z)-G(0,y)H(0,z)\Big)+G(-z,1-z,y)\nonumber\\
&&+\:G(1,0,y)-H(1,z)G(-z,y)\nonumber\\
&&\qquad+\:\frac{11}{12}\Big(-G(1-z,y)+H(1,z)-H(0,z)-G(0,y)\Big)\bigg]\nonumber\\
&&-\:\frac{z}{6(1-y-z)}\Big(1-\frac{1}{1-y-z}+\frac{z}{1-y-z}\Big)-\frac{\pi^2}{12}+i\frac{11\pi}{4}\;,\\
B^{(1)}_\beta & = & 0\;, \\
C^{(1)}_\beta & = & \frac{1}{6}\bigg[G(1-z,y)-H(1,z)+H(0,z)+G(0,y)\nonumber\\
&&-\:\frac{z}{1-y-z}\Big(-1+\frac{1}{1-y-z}-\frac{z}{1-y-z}\Big)\bigg]-i\frac{\pi}{2}\;,\\
A^{(1)}_\gamma & = & \frac{1}{6}\bigg[3\Big(-G(0,y)H(1,z)+H(0,z)G(1-z,y)+H(0,1,z)+G(0,1-z,y)\nonumber\\
&&-\:G(1,0,y)+G(1-z,0,y)\Big)+6\Big(H(1,z)G(-z,y)-G(-z,1-z,y)\Big)\nonumber\\
&&+\:5\Big(G(0,y)+H(0,z)\Big)+\frac{13}{2}\Big(-H(1,z)+G(1-z,y)\Big)-\frac{89}{6}-\frac{3(z-1)}{2y}
\bigg]\nonumber\\
&&-\:i\frac{11\pi}{4}\;,\\
B^{(1)}_\gamma & = & \frac{1}{6}\bigg[3\Big(-G(0,y)H(0,z)+G(1,0,y)-H(1,0,z)\Big)-\frac{\pi^2}{2}-\frac{27}{2}-\frac{3(z-1)}{2y}
\bigg]\;,\\
C^{(1)}_\gamma & = & \frac{1}{6}\bigg[\frac{1}{2}\Big(-G(0,y)-H(0,z)\Big)+2\Big(H(1,z)-G(1-z,y)\Big)+\frac{10}{3}\bigg]+i\frac{\pi}{2}\;.
\end{eqnarray}

\section{Two-loop helicity coefficients}
\setcounter{equation}{0}
The finite contributions to the 
renormalized two-loop helicity coefficients, decomposed 
in colour factors according to (\ref{eq:wm2}) are:

{

}

\end{appendix}


\begin{thebibliography}{99}

\bibitem{higgslep}
  R.~Barate {\it et al.} [ LEP Working Group for Higgs boson searches and ALEPH and DELPHI and L3 and OPAL Collaborations ],
  Phys.\ Lett.\  {\bf B565} (2003)  61.
  [hep-ex/0306033].


\bibitem{higgstev}
CDF and D0 Collaborations,
  [arXiv:1107.5518].


\bibitem{atlashiggs}
ATLAS Collaboration, {\it Combination of Higgs Boson 
Searches 
with up to 4.9 fb$^{-1}$ of pp Collision Data Taken at $\sqrt{s} =$ 
7 TeV with the ATLAS Experiment at the LHC},
ATLAS-CONF-2011-163.

\bibitem{cmshiggs}
CMS Collaboration, {\it Combination of CMS Searches for a
Standard Model Higgs Boson},
 CMS-PAS-HIG-11-032.


\bibitem{higgstop}
J.R.~Ellis, M.K.~Gaillard, D.V.~Nanopoulos and C.T.~Sachrajda,
Phys.\ Lett.\ B {\bf 83} (1979) 339.


\bibitem{spira}
D.~Graudenz, M.~Spira and P.M.~Zerwas,
Phys.\ Rev.\ Lett.\  {\bf 70} (1993) 1372;\\
M.~Spira, A.~Djouadi, D.~Graudenz and P.~M.~Zerwas,
Nucl.\ Phys.\ B {\bf 453} (1995) 17
[hep-ph/9504378];\\
A.~Djouadi, M.~Spira and P.M.~Zerwas,
Z.\ Phys.\ C {\bf 70} (1996) 427
[hep-ph/9511344];\\
M.~Spira,
Fortsch.\ Phys.\  {\bf 46} (1998) 203
[hep-ph/9705337].

\bibitem{effth}
F.~Wilczek,
Phys.\ Rev.\ Lett.\  {\bf 39} (1977) 1304;\\
M.A.~Shifman, A.I.~Vainshtein and V.I.~Zakharov,
Phys.\ Lett.\ B {\bf 78} (1978) 443;\\
T.~Inami, T.~Kubota and Y.~Okada,
Z.\ Phys.\ C {\bf 18} (1983) 69.

\bibitem{higgs1l}
  S.~Dawson,
  Nucl.\ Phys.\  B {\bf 359} (1991) 283.



\bibitem{higgsnnloinc}
 R.~V.~Harlander, W.~B.~Kilgore,
  Phys.\ Rev.\ Lett.\  {\bf 88} (2002)  201801
  [hep-ph/0201206];\\
C.~Anastasiou, K.~Melnikov,
  Nucl.\ Phys.\  {\bf B646} (2002)  220
  [hep-ph/0207004];
  V.~Ravindran, J.~Smith, W.~L.~van Neerven,
  Nucl.\ Phys.\  {\bf B665} (2003)  325
  [hep-ph/0302135].

\bibitem{babishiggs}
 C.~Anastasiou, K.~Melnikov, F.~Petriello,
  Nucl.\ Phys.\  {\bf B724} (2005)  197
  [hep-ph/0501130];\\
 C.~Anastasiou, G.~Dissertori, F.~Stockli,
  JHEP {\bf 0709} (2007)  018
  [arXiv:0707.2373].



\bibitem{grazzinihiggs}
  M.~Grazzini,
  JHEP {\bf 0802} (2008)  043
  [arXiv:0801.3232].

\bibitem{jetveto}
 S.~Catani, D.~de Florian, M.~Grazzini,
  JHEP {\bf 0201}, 015 (2002)
  [hep-ph/0111164].

\bibitem{t1}
 C.~F.~Berger, C.~Marcantonini, I.~W.~Stewart, F.~J.~Tackmann, W.~J.~Waalewijn,
  JHEP {\bf 1104} (2011)  092
  [arXiv:1012.4480].

\bibitem{h1j}
D.~de Florian, M.~Grazzini, Z.~Kunszt,
  Phys.\ Rev.\ Lett.\  {\bf 82} (1999)  5209
  [hep-ph/9902483];\\
 V.~Ravindran, J.~Smith, W.~L.~Van Neerven,
  Nucl.\ Phys.\  {\bf B634} (2002)  247
  [hep-ph/0201114].

\bibitem{h2j}
 J.~M.~Campbell, R.~K.~Ellis, G.~Zanderighi,
  JHEP {\bf 0610}, 028 (2006).
  [hep-ph/0608194];\\
 J.~M.~Campbell, R.~K.~Ellis, C.~Williams,
  Phys.\ Rev.\  {\bf D81} (2010)  074023.
  [arXiv:1001.4495].


\bibitem{t2}
  I.~W.~Stewart, F.~J.~Tackmann,
  [arXiv:1107.2117].


\bibitem{plehn}
  E.~Gerwick, T.~Plehn, S.~Schumann,
  [arXiv:1108.3335].



\bibitem{dixonglover}
 L.~J.~Dixon, E.~W.~N.~Glover, V.~V.~Khoze,
  JHEP {\bf 0412} (2004)  015
  [hep-th/0411092];\\
  S.~D.~Badger, E.~W.~N.~Glover and V.~V.~Khoze,
  JHEP {\bf 0503} (2005) 023
  [hep-th/0412275].


\bibitem{h1l}
S.~D.~Badger and E.~W.~N.~Glover,
  Nucl.\ Phys.\ Proc.\ Suppl.\  {\bf 160} (2006) 71
  [hep-ph/0607139].
  L.~J.~Dixon, Y.~Sofianatos,
  JHEP {\bf 0908} (2009)  058
  [arXiv:0906.0008];\\
 S.~Badger, E.~W.~N.\ Glover, P.~Mastrolia, C.~Williams,
  JHEP {\bf 1001} (2010)  036
  [arXiv:0909.4475];\\
 S.~Badger, J.~M.~Campbell, R.~K.~Ellis, C.~Williams,
  JHEP {\bf 0912} (2009)  035
  [arXiv:0910.4481].

\bibitem{schmidt}
 C.~R.~Schmidt,
  Phys.\ Lett.\  B {\bf 413} (1997) 391
  [hep-ph/9707448].

\bibitem{thanos}
  A.~Koukoutsakis,
  PhD thesis, University of Durham, 2003.


\bibitem{secdec}
T.~Binoth and G.~Heinrich,
  Nucl.\ Phys.\  B {\bf 585} (2000) 741
  [hep-ph/0004013], Nucl.\ Phys.\ B {\bf 693} (2004) 134
[hep-ph/0402265];\\
C.~Anastasiou, K.~Melnikov and F.~Petriello,
Phys.\ Rev.\ D {\bf 69} (2004) 076010
[hep-ph/0311311];\\
  G.~Heinrich,
  Int.\ J.\ Mod.\ Phys.\  A {\bf 23} (2008) 1457 
  [arXiv:0803.4177];\\
J.~Carter and G.~Heinrich,
  Comput.\ Phys.\ Commun.\  {\bf 182} (2011) 1566
  [arXiv:1011.5493];\\
 C.~Anastasiou, F.~Herzog, A.~Lazopoulos,
  JHEP {\bf 1103} (2011)  038.
  [arXiv:1011.4867]; [arXiv:1110.2368].

\bibitem{qtsub}
 S.~Catani, M.~Grazzini,
  Phys.\ Rev.\ Lett.\  {\bf 98 } (2007)  222002.
  [hep-ph/0703012].


\bibitem{antsub}
 A.~Gehrmann-De Ridder, T.~Gehrmann, E.~W.~N.~Glover,
  JHEP {\bf 0509} (2005)  056
  [hep-ph/0505111];\\
 A.~Gehrmann-De Ridder, T.~Gehrmann, E.~W.~N.~Glover, G.~Heinrich,
  JHEP {\bf 0711} (2007)  058
  [arXiv:0710.0346].


\bibitem{hadant}
 A.~Daleo, T.~Gehrmann, D.~Maitre,
  JHEP {\bf 0704 } (2007)  016
  [hep-ph/0612257];\\
 A.~Daleo, A.~Gehrmann-De Ridder, T.~Gehrmann, G.~Luisoni,
  JHEP {\bf 1001} (2010)  118
  [arXiv:0912.0374];\\
 R.~Boughezal, A.~Gehrmann-De Ridder, M.~Ritzmann,
  JHEP {\bf 1102} (2011)  098
  [arXiv:1011.6631];\\
 T.~Gehrmann, P.~F.~Monni,
  [arXiv:1107.4037].

\bibitem{joao}
  E.~W.~N.~ Glover, J.~Pires,
  JHEP {\bf 1006} (2010)  096
  [arXiv:1003.2824].


\bibitem{twol}
C.\ Anastasiou, E.W.N.~Glover, C.\ Oleari and M.E.\ Tejeda-Yeomans,
Nucl.\ Phys.\ B~{\bf 601}~(2001) 318~[hep-ph/0010212];~{\bf 601}~(2001)~347 [hep-ph/0011094];
 {\bf 605} (2001) 486 [hep-ph/0101304];\\
E.W.N.~Glover, C.~Oleari and M.E.~Tejeda-Yeomans,
Nucl.\ Phys.\ {\bf 605} (2001) 467 [hep-ph/0102201];\\
C.~Anastasiou, E.W.N.~Glover and M.E.~Tejeda-Yeomans,
Nucl.\ Phys.\ B {\bf 629} (2002) 255 [hep-ph/0201274];\\
E.W.N.~Glover and M.E.~Tejeda-Yeomans,
JHEP {\bf 0306} (2003) 033
[hep-ph/0304169];\\
E.W.N.~Glover,
JHEP {\bf 0404} (2004) 021
[hep-ph/0401119];\\
Z.~Bern, A.~De Freitas and L.J.~Dixon,
JHEP {\bf 0109} (2001) 037 [hep-ph/0109078];
JHEP {\bf 0203} (2002) 018 [hep-ph/0201161];
JHEP {\bf 0306} (2003) 028
[hep-ph/0304168];\\
A.~De Freitas and Z.~Bern,
JHEP {\bf 0409} (2004) 039
[hep-ph/0409007].

\bibitem{regge}
  V.~Del Duca and E.~W.~N.~Glover,
  JHEP\ {\bf 0110} (2001) 035
  [hep-ph/0109028];\\
 A.~V.~Bogdan, V.~Del Duca, V.~S.~Fadin and E.~W.~N.~Glover,
  JHEP\ {\bf 0203} (2002) 032
  [hep-ph/0201240].

\bibitem{3jme}
L.W.~Garland, T.~Gehrmann, E.W.N.~Glover, A.~Koukoutsakis and E.~Remiddi,
Nucl.\ Phys.\ B {\bf 627} (2002) 107 [hep-ph/0112081] and
{\bf 642} (2002) 227 [hep-ph/0206067].



\bibitem{2lsplit}
  Z.~Bern, L.~J.~Dixon and D.~A.~Kosower,
  JHEP\ {\bf 0408} (2004) 012
  [hep-ph/0404293];\\
 S.~D.~Badger and E.~W.~N.~Glover,
  JHEP {\bf 0407} (2004) 040
  [hep-ph/0405236].

\bibitem{ptdist}
R.K.~Ellis, I.~Hinchliffe, M.~Soldate, and J.J.~van der Bij,
Nucl.~Phys.~{\bf B297} (1988) 221;\\
U.~Baur and E.W.N.~Glover,
Nucl.~Phys.~{\bf B339} (1990) 38.

\bibitem{kniehl}
K.G.~Chetyrkin, B.A.~Kniehl and M.~Steinhauser,
Phys.\ Rev.\ Lett.\  {\bf 79} (1997) 353
[hep-ph/9705240].

\bibitem{kniehl2}
B.A.~Kniehl and M.~Spira,
Z.\ Phys.\ C {\bf 69} (1995) 77
[hep-ph/9505225];\\
K.G.~Chetyrkin, B.A.~Kniehl and M.~Steinhauser,
Nucl.\ Phys.\ B {\bf 510} (1998) 61
[hep-ph/9708255].

\bibitem{Xu:1986xb}
Z.~Xu, D.~H.~Zhang and L.~Chang,
Nucl.\ Phys.\ B {\bf 291} (1987) 392.



\bibitem{Berends:ie}
F.~A.~Berends, R.~Kleiss and S.~Jadach,
Nucl.\ Phys.\ B {\bf 202} (1982) 63;\\
F.~A.~Berends, R.~Kleiss, P.~de Causmaecker, R.~Gastmans, W.~Troost and T.~T.~Wu
                  [CALKUL Collaboration],
Nucl.\ Phys.\ B {\bf 239} (1984) 382, {\bf 239} (1984) 395.


\bibitem{dixon}
 L.~J.~Dixon,
Proceedings of TASI'94 ``QCD \& Beyond'', ed.\ D.\ Soper, World Scientific, 
1995, p.\ 539
  [hep-ph/9601359].


\bibitem{ourevent}
  A.~Gehrmann-De Ridder, T.~Gehrmann, E.W.N.~Glover and G.~Heinrich,
  Phys.\ Rev.\ Lett.\  {\bf 99} (2007) 132002
  [arXiv:0707.1285];
  JHEP {\bf 0712} (2007) 094
  [arXiv:0711.4711];
  Phys.\ Rev.\ Lett.\  {\bf 100} (2008) 172001
  [arXiv:0802.0813];
 JHEP {\bf 0905} (2009) 106  [arXiv:0903.4658].

\bibitem{weinzierl}
S.~Weinzierl,
 Phys.\ Rev.\ Lett.\  {\bf 101} (2008) 162001
  [arXiv:0807.3241];
  JHEP {\bf 0906} (2009) 041
  [arXiv:0904.1077];
  JHEP {\bf 0907} (2009) 009
  [arXiv:0904.1145];
  Phys.\ Rev.\  D {\bf 80} (2009) 094018
  [0909.5056].
  Eur.\ Phys.\ J.\  C {\bf 71} (2011) 1565
  [arXiv:1011.6247].

\bibitem{tancredi}
T.~Gehrmann and L.~Tancredi,
  [arXiv:1112.1531].


\bibitem{qgraf}
P.~Nogueira,
J.\ Comput.\ Phys.\  {\bf 105} (1993) 279.


\bibitem{dreg1}
C.G.\ Bollini and J.J.\ Giambiagi, Nuovo Cim.\ {\bf 12B} (1972) 20.

\bibitem{dreg2}
G.M.\ Cicuta and E.\ Montaldi, Nuovo Cim.\ Lett.\ {\bf 4} (1972) 329.

\bibitem{hv}
G.\ 't Hooft and M.\ Veltman, Nucl.\ Phys.\ {\bf B44} (1972) 189.

\bibitem{form}
J.A.M.~Vermaseren, 
{\it New features of FORM}, 
math-ph/0010025;
  Nucl.\ Phys.\ Proc.\ Suppl.\  {\bf 183} (2008)  19
  [0806.4080].



\bibitem{chet1}
F.V.\ Tkachov, Phys.\ Lett.\ {\bf 100B} (1981) 65.
\bibitem{chet2}
K.G.\ Chetyrkin and F.V.\ Tkachov, Nucl.\ Phys.\ {\bf B192} (1981) 159.

\bibitem{gr} 
T.\ Gehrmann and E.\ Remiddi, Nucl.\ Phys.\
{\bf B580} (2000) 485 [hep-ph/9912329].


\bibitem{laporta}
S.~Laporta,
Int.\ J.\ Mod.\ Phys.\ A {\bf 15} (2000) 5087
[hep-ph/0102033].

\bibitem{reduze}
 C.~Studerus,
  Comput.\ Phys.\ Commun.\  {\bf 181} (2010)  1293.
  [arXiv:0912.2546].


\bibitem{mi}
T.\ Gehrmann and E.\ Remiddi, Nucl.~Phys.~{\bf B601} (2001) 248
[hep-ph/0008287];
{\bf B601} (2001) 287 [hep-ph/0101124].



\bibitem{hpl}
E.\ Remiddi and J.A.M.\ Vermaseren, Int.\ J.\ Mod.\ Phys.\ {\bf A15}
(2000) 725 [hep-ph/9905237].

\bibitem{nielsen}
N.~Nielsen, Nova Acta Leopoldiana (Halle) {\bf 90} (1909) 123;\\
K.S.\ K\"olbig, J.A.\ Mignaco and E.\ Remiddi, BIT {\bf 10} (1970) 38.

\bibitem{hplnum}
T.~Gehrmann and E.~Remiddi,
Comput.\ Phys.\ Commun.\ {\bf 141} (2001) 296 [hep-ph/0107173];
  Comput.\ Phys.\ Commun.\  {\bf 144} (2002) 200
  [hep-ph/0111255];\\
 J.~Vollinga, S.~Weinzierl,
  Comput.\ Phys.\ Commun.\  {\bf 167} (2005)  177.
  [hep-ph/0410259];\\
 D.~Ma\^{\i}tre,
  Comput.\ Phys.\ Commun.\  {\bf 174} (2006) 222
  [hep-ph/0507152];\\
S.~B\"uhler, C.~Duhr,
  [arXiv:1106.5739].


\bibitem{ancont}
T.~Gehrmann and E.~Remiddi,
Nucl.\ Phys.\ B {\bf 640} (2002) 379
[hep-ph/0207020].

\bibitem{Harlander:2001is}
R.~V.~Harlander and W.~B.~Kilgore,
Phys.\ Rev.\ D {\bf 64}, 013015 (2001)
[hep-ph/0102241].

\bibitem{catani}
S.\ Catani, Phys.\ Lett.\ {\bf B427} (1998) 161 [hep-ph/9802439].

\bibitem{cataniproof}
G.~Sterman and M.~E.~Tejeda-Yeomans,
  Phys.\ Lett.\  B {\bf 552} (2003) 48
  [hep-ph/0210130];\\
   T.~Becher and M.~Neubert,
  Phys.\ Rev.\ Lett.\  {\bf 102} (2009) 162001
  [arXiv:0901.0722];\\
  E.~Gardi and L.~Magnea,
  JHEP {\bf 0903} (2009) 079
  [arXiv:0901.1091].


\end{thebibliography}
\end{document}